\documentclass[
  twoside,
  amsmath,
  amssymb,
  aps,
  pre,
  reprint,
  floatfix
]{revtex4-2}

\newcommand{\ArticleTitle}{%
  Logarithmic dispersion at marginal stability in collective attention%
}
\newcommand{\ArticleAuthor}{Keisuke Okamura}
\newcommand{\ArticleKeywords}{%
  attention dynamics; stochastic differential equation;
  fractional Brownian motion; memory effect%
}

\newcommand{\JournalISSN}{2470-0045}
\newcommand{\ArticleYear}{2026}
\newcommand{\ArticleVolume}{114}
\newcommand{\ArticleIssue}{5}
\newcommand{\ArticleCode}{L052302}
\newcommand{\ArticlePageCount}{5}
\newcommand{\ArticleDOI}{10.1103/c9ny-57bd}
\newcommand{\ArticleDOIURL}{https://doi.org/\ArticleDOI}

\newcommand{\JournalBanner}{%
  PHYSICAL REVIEW E \textbf{\ArticleVolume}, \ArticleCode{} (\ArticleYear)%
}
\newcommand{\ShortTitle}{LOGARITHMIC DISPERSION AT MARGINAL STABILITY \ldots}
\newcommand{\ShortAuthor}{KEISUKE OKAMURA}

\usepackage[T1]{fontenc}
\usepackage{newtxtext,newtxmath}

\usepackage{mathtools}
\usepackage{bm}

\usepackage{enumitem}

\usepackage{xcolor}
\usepackage{graphicx}
\usepackage{overpic}

\definecolor{TiffanyBlue}{cmyk}{1,0,0.3,0}
\definecolor{LetterGray}{gray}{0.50}

\usepackage{xurl}

\usepackage[
  unicode=true,
  bookmarks=true,
  colorlinks=true,
  linkcolor=blue,
  urlcolor=blue,
  citecolor=TiffanyBlue,
  plainpages=false,
  pdfpagelabels,
  breaklinks=true
]{hyperref}

\usepackage{orcidlink}

\usepackage{fancyhdr}

\makeatletter
\g@addto@macro\UrlBreaks{%
  \do\.\do\-\do\_\do\=\do\&\do\?\do\#%
}
\makeatother

\hypersetup{
  pdftitle={\ArticleTitle},
  pdfauthor={\ArticleAuthor},
  pdfkeywords={\ArticleKeywords}
}

\fancypagestyle{firstpage}{%
  \fancyhf{}%
  \fancyhead[L]{%
    \smash{%
      \raisebox{-1.35\baselineskip}[0pt][0pt]{%
        \begingroup
          \setlength{\fboxsep}{0pt}%
          \colorbox{LetterGray}{%
            \parbox[c][10.5pt][c]{90pt}{%
              \centering\color{white}\sffamily\bfseries
              \fontsize{7.5pt}{8.5pt}\selectfont Letter%
            }%
          }%
        \endgroup
      }%
    }%
  }%
  \fancyhead[C]{\normalsize \JournalBanner}%
  \fancyfoot[L]{%
    \small
    \JournalISSN/\ArticleYear/\ArticleVolume(\ArticleIssue)/%
    \ArticleCode(\ArticlePageCount)%
  }%
  \fancyfoot[C]{\small \ArticleCode-\arabic{page}}%
  \fancyfoot[R]{\small Published by the American Physical Society}%
  \renewcommand{\headrulewidth}{0.4pt}%
  \renewcommand{\footrulewidth}{0pt}%
}

\fancypagestyle{restpage}{%
  \fancyhf{}%
  \fancyhead[LE]{\small \ShortAuthor}%
  \fancyhead[RE]{\small \JournalBanner}%
  \fancyhead[LO]{\small \ShortTitle}%
  \fancyhead[RO]{\small \JournalBanner}%
  \fancyfoot[C]{\small \ArticleCode-\arabic{page}}%
  \renewcommand{\headrulewidth}{0.4pt}%
  \renewcommand{\footrulewidth}{0pt}%
}

\fancypagestyle{suppl1}{%
  \fancyhf{}%
  \fancyhead[C]{\normalsize \JournalBanner}%
  \fancyfoot[C]{\small \ArticleCode-suppl-\arabic{page}}%
  \renewcommand{\headrulewidth}{0pt}%
  \renewcommand{\footrulewidth}{0pt}%
}

\fancypagestyle{suppl2}{%
  \fancyhf{}%
  \fancyfoot[C]{\small \ArticleCode-suppl-\arabic{page}}%
  \renewcommand{\headrulewidth}{0pt}%
  \renewcommand{\footrulewidth}{0pt}%
}

\makeatletter
\let\RRAP@orig\frontmatter@RRAPformat
\def\frontmatter@RRAPformat#1{%
  \vskip 6pt\relax
  \hrule height\z@\relax
  \RRAP@orig{#1}%
}
\makeatother

\newlength{\captionindent}
\makeatletter
\let\CAP@fnumfigure\fnum@figure
\let\CAP@fnumtable\fnum@table
\renewcommand{\fnum@figure}{\hspace*{\captionindent}\CAP@fnumfigure}
\renewcommand{\fnum@table}{\hspace*{\captionindent}\CAP@fnumtable}
\makeatother

\newcommand{\runin}[1]{\textit{#1}\ \ignorespaces}
\let\SIrunin\runin

\newcommand{\dd}[1]{\mathop{}\!\mathrm{d}#1}
\newcommand{\mean}[1]{\left\langle #1\right\rangle}
\newcommand{\Var}{\operatorname{Var}}
\newcommand{\Cov}{\operatorname{Cov}}
\newcommand{\tc}{t_{\mathrm{c}}}

\newcommand{\safegraphic}[2][]{%
  \IfFileExists{fig/#2}{%
    \includegraphics[#1]{fig/#2}%
  }{%
    \fbox{%
      \parbox[c][0.28\textheight][c]{0.88\linewidth}{%
        \centering
        Missing figure file:\\[0.5em]
        \texttt{\detokenize{fig/#2}}%
      }%
    }%
  }%
}

\newcommand{\panel}[3][]{%
  \IfFileExists{fig/#3}{%
    \begin{overpic}[width=0.32\textwidth,percent,#1]{fig/#3}%
      \put(1.6,70){\makebox(0,0)[lt]{#2}}%
    \end{overpic}%
  }{%
    \fbox{%
      \parbox[c][0.20\textwidth][c]{0.29\textwidth}{%
        \centering
        #2\\[0.5em]
        \texttt{\detokenize{fig/#3}}%
      }%
    }%
  }%
}

\begin{document}

\title{\vspace*{16pt}\ArticleTitle}

\author{\ArticleAuthor\,\orcidlink{0000-0002-0988-6392}}
\email[Contact author:~]{okamura@alumni.lse.ac.uk}
\affiliation{Ministry of Education, Culture, Sports, Science and Technology, Tokyo, Japan}

\received{18 April 2026}
\accepted[accepted ]{23 October 2026}
\published[published ]{12 November 2026}

\begin{abstract}
How cross-sectional heterogeneity evolves in nonequilibrium populations of competing, ageing units is a collective question that cannot be resolved from a representative trajectory or the ensemble mean alone.
Analysis of more than two million age-aligned English Wikipedia page-view trajectories from six yearly cohorts spanning 2015--2020 shows that the variance of logarithmic attention grows linearly with $\ln t$, while the ensemble mean relaxes on the same logarithmic clock.
This dispersion law recurs across all six cohorts and is systematically favoured over the tested power-law alternative; temporal increments also exhibit consistent antipersistent scaling.
A complementary trajectory-level analysis reveals broad, non-Gaussian variation in effective relaxation rates, showing that this collective regularity coexists with heterogeneous observed trajectories.
A minimal scale-free stochastic model combines mean relaxation with self-affine temporal fluctuations of age-dependent amplitude.
Within this model class, logarithmic spreading emerges when fluctuation accumulation balances ageing attenuation, defining the marginal boundary between algebraic broadening and saturation.
Together, age-resolved cross-sectional dispersion and temporal-increment scaling provide a transferable ensemble-level criterion for distinguishing amplification, marginal spreading, and saturation in other populations of competing, ageing units.
\par\vspace{0.75\baselineskip}\noindent
DOI: \href{\ArticleDOIURL}{\ArticleDOI}
\end{abstract}

\keywords{\ArticleKeywords}

\maketitle

\thispagestyle{firstpage}
\pagestyle{restpage}

\runin{Introduction.}%
In many nonequilibrium systems, individual trajectories and their ensemble average provide the conventional starting point for describing populations of competing units.
They characterise unit-level evolution and mean behaviour, whereas ensemble dispersion---quantified here by the variance---captures how heterogeneity develops across the population.
Yet the scaling of this dispersion with age has received comparatively little attention, despite its potential to distinguish qualitatively different collective regimes of amplification, marginal spreading, and saturation.

Collective attention provides a particularly well-resolved empirical setting for this question: the activity of many items competing for visibility can be followed simultaneously over time.
Across information environments, collective attention exhibits bursts, cascade structures, and rapid turnover \cite{Barabasi05,Goel16,Lorenz-Spreen19}.
At the level of individual items, existing descriptions include stochastic relaxation and novelty decay \cite{Dezso06,Wu07}, self-exciting or epidemic cascades \cite{Hawkes71,Crane08}, and preferential or multiplicative growth mechanisms \cite{Barabasi99,Ratkiewicz10}.
Models with explicit competition among multiple items further explain heterogeneity and critical popularity statistics under limited attention \cite{Gleeson14,Weng12}.
These approaches, however, do not determine how cross-sectional dispersion evolves with age.
Consequently, trajectory-level relaxation laws or static popularity statistics alone do not locate the marginal boundary between amplification and saturation considered here.

Here we address this problem using English Wikipedia page views, a large-scale observational record of collective attention across a vast information space.
Across six yearly cohorts from 2015 to 2020, we find that the ensemble mean and variance consistently evolve on the same logarithmic clock: the variance of logarithmic attention grows linearly with $\ln t$, while its mean relaxes proportionally to $-\ln t$.
Temporal increments in all six cohorts also exhibit consistent antipersistent scaling.
For the 2020 cohort examined at the trajectory level, finite-trajectory estimates of effective decay rates form a broad, non-Gaussian distribution.
Together, these observations show that reproducible ensemble-level scaling coexists with pronounced heterogeneity among observed trajectories.

To account for these ensemble laws, we formulate a minimal scale-free stochastic model for logarithmic attention that combines self-affine temporal correlations with an age-dependent fluctuation amplitude.
Using fractional Brownian motion (fBM) as a parsimonious Gaussian reference for the measured second-order correlations \cite{Mandelbrot68}, the model introduces ageing attenuation into the constant-amplitude setting previously considered for citation dynamics \cite{Okamura25}.
Within this model class, logarithmic dispersion emerges when self-affine fluctuation accumulation balances ageing attenuation, marking the boundary between algebraic broadening and saturation.
The resulting description locates marginality at the level of ensemble dispersion rather than representative article dynamics and provides a transferable diagnostic, based on age-resolved dispersion and temporal-increment scaling, for other populations in which activity is allocated competitively.

\begin{figure*}[t!]
\centering
\begin{tabular*}{\textwidth}{@{\extracolsep{\fill}}ccc@{}}
\panel{(a)}{week_period_2020.pdf} &
\panel{(b)}{est_H_2020.pdf} &
\panel{(c)}{est_gamma_2020.pdf} \\[0.2em]
\panel{(d)}{est_a_2020.pdf} &
\panel{(e)}{est_params.pdf} &
\panel{(f)}{phase_diagram.pdf}
\end{tabular*}
\caption{Empirical scaling and the scale-free phase portrait of collective attention.
(a) Median post-selection trajectory across the 2020 trajectory episodes, showing slow decay with weekly modulation.
(b) Pooled increment-variance scaling
$\Var[Y_{t+\tau}-Y_{t}]\propto\tau^{2H}$,
yielding $H\approx0.22$.
(c) Ensemble variance as a function of $\ln t$; the fitted linear regime gives
$\Var[Y_{t}]=\gamma\ln t+\mathrm{const}$.
(d) Ensemble mean relative to the selection-day value; its linear dependence on $\ln t$ gives
$\mean{Y_{t}-Y_{0}}=-a\ln t+\mathrm{const}$.
(e) Estimated $H$, $\gamma$, and $a$ across the 2015--2020 cohorts.
(f) Phase diagram of the scale-free model introduced below.
For a fluctuation amplitude $\beta(t)\propto t^{-\eta}$, the coordinate $\xi=H-\eta$ separates algebraic amplification ($\xi>0$), marginal logarithmic spreading ($\xi=0$), and saturation ($\xi<0$).
Dotted vertical lines in (c) and (d) mark the fitted endpoint $t=\tc$.}
\label{fig:summary}
\end{figure*}

\runin{Data and empirical scaling.}%
To construct the age-aligned ensembles analysed here, we combine the English Wikipedia `Topviews' rankings with page-view records for 2015--2020 \cite{WikimediaTopviews,WikimediaPageviews}.
For each reference day $d$, we select the 1,000 most-viewed articles and follow their activity over the subsequent $T=365$ days.
Because the Pageviews API record begins on July 1, reference days cover July--December in 2015 and the full calendar year thereafter.
Further data-construction details and cohort sizes are provided in the Supplemental Material (SM) \cite[Sec.~\ref{SI:data_api} and Table~\ref{tab:cohort_counts}]{suppletext}.

For a trajectory episode, let $V_{t}$ denote the page views at elapsed day $t$, with $V_{0}$ denoting the selection-day count, and define
\begin{equation}\label{eq:Y_data}
X_{t}=V_{t}+1,\quad Y_{t}=\ln X_{t}.
\end{equation}
The unit shift retains zero-view days while having negligible effect when page-view counts are large.
Each reference-day--article pair is treated as a distinct trajectory episode, so repeated appearances of the same article represent distinct episodes of salience.
All moment curves are calculated across these trajectory episodes at fixed elapsed time.
The resulting ensemble laws therefore characterise the post-selection evolution of salient articles rather than pre-selection dynamics or the full population of Wikipedia pages.

Figure~\ref{fig:summary}(a) shows the typical post-selection relaxation together with weekly modulation.
We estimate temporal roughness from increments, which reduce the influence of the slowly varying mean.
For each dyadic lag $\tau\in\{1,2,4,8,16,32\}$, increments are pooled across trajectory episodes and all admissible post-selection elapsed times within each yearly cohort.
Their empirical variance is well described by the cohort-level lag scaling
\begin{equation}\label{eq:scaling_H}
\Var[Y_{t+\tau}-Y_{t}]\propto\tau^{2H}.
\end{equation}
For 2020, the fit gives $\hat{H}\approx0.22$ [Fig.~\ref{fig:summary}(b)]; across all six cohorts, the estimates lie in the narrow range $\hat{H}\approx0.21$--$0.23$ [Fig.~\ref{fig:summary}(e) and SM \cite[Sec.~\ref{SI:est_H} and Fig.~\ref{fig:est_H_aggr}]{suppletext}].
Within an fBM reference description, the sublinear exponent $2\hat{H}<1$ corresponds to antipersistent fluctuations and sample paths rougher than Brownian motion.
Equation~\eqref{eq:scaling_H} should be read as an effective finite-window lag scaling; its geometric interpretation and relation to the local increment law under an age-dependent fluctuation amplitude are detailed in the SM \cite[Secs.~\ref{SI:scaling_H} and \ref{SI:geometric}]{suppletext}.

The principal empirical observable is the cross-sectional variance of $Y_{t}$.
As shown for 2020 in Fig.~\ref{fig:summary}(c), it grows linearly with logarithmic age,
\begin{equation}\label{eq:empirical_variance}
\Var[Y_{t}]=\gamma\ln t+\mathrm{const},
\end{equation}
over the data-adaptive range $1\leq t\leq\tc$.
The endpoint $\tc$ is determined by a single-breakpoint least-squares analysis of the variance curve and then held fixed for the mean fit and the power-law comparison \cite[Sec.~\ref{SI:tc_fit}]{suppletext}.
A direct comparison with a two-parameter power law over the same fitted interval favours the logarithmic form in every yearly cohort \cite[Sec.~\ref{SI:compare_powerlaw} and Table~\ref{tab:model_comparison}]{suppletext}.
The corresponding cohort-wise variance curves are shown in the SM \cite[Fig.~\ref{fig:est_gamma_aggr}]{suppletext}.

The mean evolves on the same logarithmic time axis [Fig.~\ref{fig:summary}(d)]:
\begin{equation}\label{eq:scaling_a}
\mean{Y_{t}-Y_{0}}=-a\ln t+\mathrm{const}.
\end{equation}
The additive constant absorbs the difference between the selection-day value and the first post-selection observation used in the scaling fit.
Equation~\eqref{eq:scaling_a} corresponds to an effective cohort-level drift proportional to $-1/t$.
Across years, $\hat{H}$ and $\hat{\gamma}$ remain comparatively stable, whereas $\hat{a}$ shows a modest overall upward tendency, increasing from $0.216$ in 2015 to $0.277$ in 2020 despite nonmonotonic year-to-year variation (Fig.~\ref{fig:summary}(e) and SM \cite[Sec.~\ref{SI:cross_year_estimates}, Fig.~\ref{fig:est_a_aggr}, and Table~\ref{tab:summary}]{suppletext}).
Larger values of $a$ correspond to faster mean relaxation, so this overall tendency indicates a steeper decline in the later cohorts.

The coefficients $a$, $\gamma$, and $H$ describe complementary aspects of the dynamics: $a$ governs relaxation of the distribution centre, $\gamma$ measures the logarithmic growth rate of cross-sectional variance, and $H$ characterises temporal-increment scaling.
Together, the common logarithmic clock of the first two ensemble moments and the cross-year stability of $H$ and $\gamma$ define a reproducible ensemble structure even as the mean-relaxation rate varies.

\runin{Scale-free stochastic model.}%
The logarithmic variance growth in Eq.~\eqref{eq:empirical_variance} is the central empirical clue.
The observed growth is slower than the algebraic broadening generated by unattenuated self-affine fluctuations, yet shows no flattening towards saturation over the fitted window.
This intermediate behaviour motivates a scale-free stochastic model for logarithmic attention.

Let $X(t)>0$ denote the continuous-time model counterpart of the empirical positive variable $X_{t}=V_{t}+1$ defined in Eq.~\eqref{eq:Y_data}, and set $Y(t)\coloneq\ln X(t)$.
We write its evolution as
\begin{equation}\label{eq:log_model}
Y(t)-Y(t_{0})=\int_{t_{0}}^{t}\alpha(s)\dd{s}+\int_{t_{0}}^{t}\beta(s)\dd{B_{H}(s)},\quad t\geq t_{0}>0.
\end{equation}
Here $B_{H}$ is fBM with Hurst exponent $H\in(0,1)$ \cite{Mandelbrot68}, $\alpha(t)$ is a deterministic drift, and $\beta(t)$ is an age-dependent fluctuation amplitude.
We identify $t_{0}$ with the first post-selection day entering the scaling analysis; although $t_{0}=1$ day empirically, we retain it symbolically to keep dimensionless time ratios and scale transformations explicit.
The stochastic construction and its relation to the positive process $X(t)$ are detailed in the SM \cite[Sec.~\ref{SI:stochastic_calculus}]{suppletext}.

The empirical logarithmic relaxation of the mean corresponds to $\alpha(t)=-a/t$, while the effective increment scaling in Eq.~\eqref{eq:scaling_H} provides the estimate of the self-affine exponent $H$ used below.
To allow the fluctuations to weaken with age without introducing a characteristic time, we use the minimal scale-free ansatz
\begin{equation}\label{eq:beta_power}
\beta(t)=\frac{b}{t_{0}^{H}}\left(\frac{t}{t_{0}}\right)^{-\eta},\quad \eta\geq0,
\end{equation}
where $b>0$ sets the overall fluctuation scale; the factor $t_{0}^{-H}$ fixes its normalisation relative to the reference time.
Equation~\eqref{eq:beta_power} provides a phenomenological interpolation between unattenuated fBM fluctuations at $\eta=0$ and progressively stronger attenuation at larger $\eta$.
Writing
\begin{equation}\label{eq:I_def}
I(t;t_{0})\coloneq\int_{t_{0}}^{t}\beta(s)\dd{B_{H}(s)},
\end{equation}
the deterministic drift gives
\begin{equation}
\Var[Y(t)-Y(t_{0})]=\Var[I(t;t_{0})].
\end{equation}
Under the simplifying ensemble assumption that $Y(t_{0})$ is uncorrelated with the subsequent stochastic accumulation, the initial cross-sectional variance contributes an additive baseline to $\Var[Y(t)]$; the general covariance relation is given in the SM \cite[Sec.~\ref{SI:stochastic_calculus}]{suppletext}.

The competition between fluctuation accumulation and ageing attenuation becomes transparent by comparing intervals of fixed logarithmic length $[t,\rho t]$, with $\rho>1$.
The stochastic contribution over such an interval is the increment
\begin{equation}\label{eq:log_interval_contr}
I_{\rho}(t)\coloneq I(\rho t;t_{0})-I(t;t_{0})
=\frac{b}{t_{0}^{H}}\int_{t}^{\rho t}\left(\frac{s}{t_{0}}\right)^{-\eta}\dd{B_{H}(s)}.
\end{equation}
Self-similarity of $B_H$ supplies a factor $t^{2H}$ to the variance, whereas the homogeneous amplitude supplies a factor $t^{-2\eta}$.
Consequently,
\begin{equation}
\Var[I_{\rho}(t)]=\left(\frac{t}{t_{0}}\right)^{2(H-\eta)}\Var[I_{\rho}(t_{0})].
\end{equation}
The net exponent governing the magnitude of fluctuations accumulated per logarithmic time interval is therefore
\begin{equation}\label{eq:xi}
\xi\coloneq H-\eta.
\end{equation}
Equivalently, under the dilation $t\mapsto\rho t$ the root-mean-square fluctuation accumulated over a fixed logarithmic interval rescales by $\rho^{\xi}$; $\xi=0$ is marginal in the precise sense that this fluctuation scale is invariant under dilations of age.
Thus $\xi$ serves as the phase coordinate distinguishing the accumulation regimes.

A logarithmic-time decomposition, including correlations between successive intervals, yields \cite[Sec.~\ref{SI:scaling_log}]{suppletext}
\begin{equation}\label{eq:variance_regimes}
\Var[I(t;t_{0})]\sim
\begin{cases}
~C_{+}(t/t_{0})^{2\xi},&\xi>0,\\[2pt]
~C_{0}\ln(t/t_{0}),&\xi=0,\\[2pt]
~C_{-},&\xi<0,
\end{cases}
\end{equation}
where $C_{+}$, $C_{0}$, and $C_{-}$ are positive constants.
For $\xi>0$, successive logarithmic time intervals contribute fluctuations of increasing magnitude, producing algebraic broadening.
For $\xi<0$, their contributions weaken with age and the accumulated variance saturates.
At $\xi=0$, intervals related by a fixed rescaling contribute on the same scale.
The corresponding increments then form a stationary sequence in logarithmic time with summable correlations, and their cumulative variance gives the ultraslow $\ln(t/t_{0})$ law \cite[Sec.~\ref{SI:scaling_log}]{suppletext}.
The logarithmic law is therefore not postulated separately, but emerges as the marginal member of the same scale-free family that produces algebraic broadening for $\xi>0$ and saturation for $\xi<0$.
Under the ensemble assumption stated above, these are also the asymptotic regimes of the age-dependent part of $\Var[Y(t)]$, with the initial variance absorbed into the additive constant in Eq.~\eqref{eq:empirical_variance}.

The two empirical measurements play complementary roles: temporal-increment scaling estimates $H$, whereas the evolution of cross-sectional variance identifies the accumulation regime [Figs.~\ref{fig:summary}(b) and \ref{fig:summary}(c)].
Within this ansatz, the logarithmic variance places the ensembles near $\xi=0$; combining this result with the measured $H$ then gives $\eta\simeq H$ without imposing the balance a priori.
Across cohorts, the variance curves are consistent with marginality over the fitted windows, showing neither the systematic upward curvature predicted for $\xi>0$ nor the flattening predicted for $\xi<0$ \cite[Sec.~\ref{SI:log_slope} and Table~\ref{tab:SM_slope_signatures}]{suppletext}.

The resulting phase portrait [Fig.~\ref{fig:summary}(f)] is dynamically reminiscent of spreading and arrest in glassy systems \cite{Bouchaud92}: `liquid-like' amplification for $\xi>0$ and `frozen' saturation for $\xi<0$ are separated by ultraslow spreading at the marginal boundary $\xi=0$.
These labels refer to contrasting regimes of ensemble-dispersion dynamics rather than thermodynamic phases and provide a structural language for comparing amplification, marginal spreading, and saturation across nonequilibrium ageing systems.

At marginality, setting $\eta=H$ cancels the reference-time factors in Eq.~\eqref{eq:beta_power} and gives the compact model
\begin{equation}\label{eq:marginal_model}
\dd{Y(t)}=-\frac{a}{t}\dd{t}+\frac{b}{t^{H}}\dd{B_{H}(t)}.
\end{equation}
Equation~\eqref{eq:marginal_model} is shorthand for the canonical Wiener-integral construction in Eq.~\eqref{eq:log_model}, as detailed in the SM \cite[Sec.~\ref{SI:stochastic_calculus}]{suppletext}.
The same empirically estimated exponent $H$ thus characterises the self-affine driver and the ageing of its amplitude.
Within this marginal ensemble description, the empirical slope $\gamma$ corresponds to the coefficient $C_{0}$ in Eq.~\eqref{eq:variance_regimes}.
At fixed $H$ and fBM normalisation, $C_{0}\propto b^{2}$; hence, once $H$ is determined from increment scaling, the measured $\gamma$ calibrates the fluctuation scale $b$ \cite[Sec.~\ref{SI:scaling_log}]{suppletext}.

\runin{Summary and discussion.}%
The central result is a reproducible logarithmic law for cross-sectional dispersion in collective attention, observed across all six yearly cohorts alongside logarithmic mean relaxation and stable antipersistent increment scaling.
Within the scale-free model, this dispersion law marks the marginal balance between self-affine fluctuation accumulation and the age-dependent attenuation of fluctuation amplitude.
At this balance, each fixed multiplicative interval in age contributes fluctuations of the same order, so that the variance grows in proportion to the number of such intervals and hence logarithmically with age; weaker or stronger attenuation instead produces algebraic broadening or saturation, respectively.
Because both the mean drift and the fluctuation amplitude decrease algebraically with elapsed age, the effective pace of evolution slows as an article ages.
This explicit age dependence provides a concrete connection to ageing dynamics and anomalous transport \cite{Metzler00}.

Although Eq.~\eqref{eq:marginal_model} is written at the trajectory level, it is calibrated to the observed first- and second-order ensemble laws and should not be read as a complete generative model for every article.
The distribution of fitted trajectory-level relaxation coefficients is broad and markedly non-Gaussian
\cite[Secs.~\ref{SI:gmm}--\ref{SI:gmm_scope} and Fig.~\ref{fig:gmm_density_2020}]{suppletext}.
This pronounced heterogeneity coexists with cross-year stable values of $H$ and $\gamma$, showing that reproducible second-order scaling does not require trajectory-level uniformity.
The logarithmic law thus expresses collective statistical organisation across heterogeneous attention histories: marginality characterises the evolution of ensemble dispersion, not individual articles.

The cross-cohort comparison further separates mean relaxation from dispersion dynamics.
The estimates of $H$ and $\gamma$ remain comparatively stable, whereas the coefficient $a$ exhibits a modest overall upward tendency.
Because $a$ enters only through the cohort-level drift $\alpha(t)=-a/t$, its variation changes the relaxation of the distribution centre without, by itself, changing the accumulation coordinate $\xi$.
The larger values of $a$ in the later cohorts are consistent with the acceleration of collective attention dynamics reported across information environments \cite{Lorenz-Spreen19}, although the nonmonotonic sequence of six cohorts is insufficient to establish a secular trend or identify its cause.

The comparison with citation dynamics clarifies the role of the age-dependent fluctuation amplitude.
In Ref.~\cite{Okamura25}, a constant amplitude allowed self-affine temporal correlations to produce algebraic growth of the cross-sectional variance of logarithmic citation counts.
Here, scale-free attenuation balances self-affine accumulation, yielding logarithmic growth of the cross-sectional variance of log-transformed page views.
Thus the temporal correlation exponent alone does not determine whether dispersion broadens algebraically, grows logarithmically, or saturates; the age dependence of the fluctuation amplitude is equally decisive.

What selects the apparent balance $\eta\simeq H$ remains open.
Competition for finite collective attention may contribute, but the present observations do not distinguish external resource constraints from platform-mediated exposure, endogenous feedback, or other mechanisms.
A microscopic derivation of Eq.~\eqref{eq:marginal_model} would therefore need to account for both the correlated effective noise and the ageing of its amplitude.

Several qualifications delimit the present conclusion.
The analysis concerns post-selection episodes of salient articles in English Wikipedia rather than the full population of pages, and the six cohorts establish temporal reproducibility within one information environment rather than replication across independent systems.
The finite fitted windows support a robust logarithmic regime but do not establish an exact asymptotic law.
Moreover, fBM is used as a parsimonious Gaussian reference for the measured second-order correlations; neither the event-level origin of this effective process nor the microscopic source of trajectory-level heterogeneity is uniquely identified.
In particular, the distribution of fitted relaxation rates cannot be equated directly with a distribution of latent article parameters.

These limitations define the natural empirical tests of the framework.
The analysis requires age-resolved activity trajectories but not user-level microdata.
Longer observation windows, other Wikipedia language editions, other information platforms, and other populations of competing, ageing units can determine whether logarithmic spreading persists or crosses over to algebraic broadening or saturation.
Joint measurement of temporal-increment scaling and cross-sectional dispersion provides a common basis for these comparisons even when individual trajectory shapes differ substantially.
The broader implication is that the scaling of ensemble dispersion supplies a collective coordinate for locating a stability boundary whose position is not determined by a representative trajectory or the ensemble mean alone.

\begin{acknowledgments}
This research received no specific grant from any funding agency in the public, commercial, or not-for-profit sectors.
The views and conclusions expressed herein are solely those of the author and should not be construed as necessarily reflecting the official policies or endorsements---whether explicit or implicit---of any organisations with which the author is presently or has previously been affiliated.
\end{acknowledgments}

\runin{Data availability.}%
The data that support the findings of this article are openly available \cite{OkamuraData2026}.


%



\onecolumngrid
\clearpage

\setcounter{page}{1}
\setcounter{section}{0}
\setcounter{subsection}{0}
\setcounter{equation}{0}
\setcounter{figure}{0}
\setcounter{table}{0}

\renewcommand{\thepage}{S\arabic{page}}
\renewcommand{\thesection}{S\arabic{section}}
\renewcommand{\thesubsection}{S\arabic{section}\Alph{subsection}}
\renewcommand{\theequation}{S\arabic{equation}}
\renewcommand{\thefigure}{S\arabic{figure}}
\renewcommand{\thetable}{S\arabic{table}}

\providecommand*{\theHsection}{}
\providecommand*{\theHsubsection}{}
\providecommand*{\theHequation}{}
\providecommand*{\theHfigure}{}
\providecommand*{\theHtable}{}

\renewcommand*{\theHsection}{SM.section.\arabic{section}}
\renewcommand*{\theHsubsection}{%
  SM.subsection.\arabic{section}.\arabic{subsection}%
}
\renewcommand*{\theHequation}{SM.equation.\arabic{equation}}
\renewcommand*{\theHfigure}{SM.figure.\arabic{figure}}
\renewcommand*{\theHtable}{SM.table.\arabic{table}}

\setcounter{secnumdepth}{2}

\makeatletter

\renewcommand{\p@subsection}{}

\renewcommand{\subsection}{%
  \@startsection
    {subsection}%
    {2}%
    {\z@}%
    {.8cm \@plus1ex \@minus .2ex}%
    {.5cm}%
    {\normalfont\small\bfseries\raggedright}%
}

\makeatother

\thispagestyle{suppl1}
\pagestyle{suppl2}

\fancypagestyle{plain}{%
  \fancyhf{}%
  \fancyfoot[C]{\small \ArticleCode-suppl-\arabic{page}}%
  \renewcommand{\headrulewidth}{0pt}%
  \renewcommand{\footrulewidth}{0pt}%
}

\begin{center}
  {\fontsize{13pt}{14pt}\selectfont\bfseries Supplemental Material}\\[1em]
  {\large for `\ArticleTitle'}\\[1em]
\end{center}

\vspace{2em}

\noindent
This Supplemental Material accompanies the article published in \textit{Phys.~Rev.~E} (DOI: \href{\ArticleDOIURL}{\ArticleDOI}) and contains:

\begin{itemize}[
  leftmargin=2em,
  itemsep=0.2em,
  topsep=0.5em,
  parsep=0pt,
  partopsep=0pt
]
  \item Methods (Sec.~\ref{SI:methods})
  \item Theory and Diagnostics (Sec.~\ref{SI:theory})
  \item Trajectory Heterogeneity (Sec.~\ref{SI:analysis})
  \item Supplemental Figures
        \ref{fig:est_H_aggr}--\ref{fig:gmm_bic_2020}
  \item Supplemental Tables
        \ref{tab:cohort_counts}--\ref{tab:GMM}
\end{itemize}

\vspace{2em}

\twocolumngrid

\section{Methods}\label{SI:methods}
This section documents the construction of the age-aligned ensembles, the estimation of the empirical scaling laws, and the associated fitting-range, diagnostic, and model-comparison procedures.
It also specifies the conditional scope of the reported regression uncertainties and the limitations arising from repeated article appearances and temporally dependent observations.

\subsection{Data acquisition via Topviews and AQS}\label{SI:data_api}
Daily page-view trajectories were obtained by combining the English Wikipedia `Topviews' rankings with the Analytics Query Service (AQS) Pageviews API \cite{WikimediaTopviews,WikimediaPageviews}.
The processed datasets supporting the findings of this study are available at Zenodo \cite{OkamuraData2026}.
For each reference day $d$, we retrieved the 1,000 most-viewed articles listed by Topviews and collected their daily page views over the subsequent common horizon of $T=365$ days.
The reference days cover July 1--December 31 in 2015 and the complete calendar year in each cohort from 2016 through 2020.
The resulting numbers of usable trajectory episodes and post-selection observations are reported in Table~\ref{tab:cohort_counts}.

\SIrunin{Selection-day alignment.}
For each reference day $d$, the Topviews endpoint provides the ranking and the associated selection-day view count $V_{0}$.
The ranking determines inclusion in the reference-day cohort, while $V_{0}$ supplies the empirical reference level used in the mean-relaxation analysis.
The Pageviews API is queried over the interval from $d+1$ through $d+T$, yielding the aligned post-selection trajectory
$\{V_{t}\}_{t=1}^{T}$.
The second-order statistics are calculated from this post-selection trajectory using the actual elapsed-day labels rather than array indices.
This convention separates cohort selection from subsequent evolution and prevents the selection-conditioned observation at $t=0$ from entering the fitted second-order scaling laws.

\SIrunin{Title handling and trajectory units.}
Each title string returned by Topviews was submitted to AQS after URL encoding where required.
Each $\{$reference day $d$, title$\}$ pair was treated as one trajectory episode.
The same article may therefore appear on multiple reference days, with each occurrence representing a distinct episode of salience.
The empirical moment curves are descriptive cross-sectional moments across these episodes at fixed elapsed time and do not require the episodes to be statistically independent.
The regression standard errors reported for the fitted moment curves quantify fit uncertainty conditional on the aggregated curves.
They should not be interpreted as cluster-robust sampling uncertainties for a population of independent articles.

\SIrunin{API parameters and missing records.}
All AQS queries used \path{project=en.wikipedia.org}, \path{access=all-access}, and \path{agent=user}.
Requests affected by rate limits or transient failures were retried with exponential backoff.
A returned page-view count of zero was retained as a valid observation and subsequently included through the transformation $Y_{t}=\ln(V_{t}+1)$.
Persistently unavailable or malformed responses were treated as missing, and only trajectory episodes with complete finite post-selection records were retained in the aligned yearly arrays.
Accordingly, the number of daily observations in Table~\ref{tab:cohort_counts} is exactly $365$ times the number of retained trajectory episodes.
The close agreement of the effective Hurst estimates across cohorts of different sizes shows that they are not visibly driven by cohort size.
It does not, however, exclude bias from systematically nonrandom missingness.

\begin{table}[t!]
\centering
\caption{Numbers of trajectory episodes and aligned post-selection observations.
For each reference-year cohort, the table lists the number of retained
$\{$reference day, title$\}$ trajectory episodes and the corresponding number of daily observations over the common horizon $T=365$.
The observation counts are exactly $365$ times the trajectory counts.
The smaller 2015 cohort reflects the July 1 start of the Pageviews API record; each retained episode nevertheless contains the full 365-day post-selection trajectory.}
\label{tab:cohort_counts}
\vspace{1em}
\begin{ruledtabular}
\begin{tabular}{c@{\hspace{2.2em}}r@{\hspace{2.2em}}r}
\raisebox{0.2em}{Reference year} & \raisebox{0.2em}{Trajectory episodes} & \raisebox{0.2em}{Daily observations} \\
\colrule
2015 & 182,971 &  66,784,415 \\
2016 & 361,908 & 132,096,420 \\
2017 & 364,941 & 133,203,465 \\
2018 & 364,948 & 133,206,020 \\
2019 & 364,945 & 133,204,925 \\
2020 & 365,930 & 133,564,450 \\
\end{tabular}
\end{ruledtabular}
\end{table}

\SIrunin{Temporal coverage of the 2015 cohort.}
The AQS Pageviews record begins on July 1, 2015, so the 2015 reference-day cohort covers only the second half of that calendar year.
Each retained 2015 episode nevertheless has the same post-selection horizon $T=365$ as the later cohorts.
The observable definitions and estimation procedures are otherwise identical across years.
The agreement of the 2015 estimates with those from the five full-calendar-year reference cohorts provides a descriptive check that the shorter selection period does not determine the reported scaling behaviour.

\subsection{Estimation of the effective Hurst exponent}\label{SI:est_H}
Hurst exponents are commonly estimated using detrended fluctuation analysis (DFA) \cite{Peng94}.
Reliable DFA estimation requires a sufficiently broad scaling range and careful treatment of nonstationary trends \cite{Hu01,Bardet08}.
In particular, deterministic and periodic trends can generate apparent crossovers in the estimated scaling law \cite{Hu01}.
The present trajectories have a limited horizon ($T=365$), a slowly relaxing mean, and visible weekly modulation.
Moreover, the reduced model contains both an age-dependent drift and an age-dependent fluctuation amplitude, so polynomial detrending would not directly restore stationary fluctuations.
We therefore estimate the second-order lag scaling directly from increments, avoiding the choice of a detrending polynomial and reducing the contribution of the slowly varying mean.

For the fractional Brownian motion (fBM) reference process with Hurst exponent $H$, increments of lag $\tau$ obey the second-order scaling
\begin{equation}\label{eq:S_H_scaling}
\Var[B_{H}(t+\tau)-B_{H}(t)]=\tau^{2H}.
\end{equation}
In the age-dependent model developed in the Letter, the corresponding relation for logarithmic attention is local: for $t\gg\tau$, age enters the leading prefactor of the increment variance, whereas the local lag exponent remains $2H$ (Sec.~\ref{SI:scaling_H}).
We use pooled empirical increments to obtain an effective estimate of this lag-scaling exponent.
Let $Y_{k,t}$ denote the logarithmic page views of trajectory episode $k$ at elapsed day $t$, and define
\begin{equation}
\Delta_{\tau}Y_{k,t}\coloneq Y_{k,t+\tau}-Y_{k,t}.
\end{equation}
For each yearly cohort and each dyadic lag
$\tau\in\{1,2,4,8,16,32\}$, we calculate
\begin{equation}\label{eq:SM_pooled_increment_variance}
S_{2}^{\mathrm{pool}}(\tau)\coloneq\Var_{k,t}\!\left[\Delta_{\tau}Y_{k,t}\right],
\end{equation}
where the variance is taken jointly over all retained trajectory episodes $k$ and all admissible elapsed days $t\in\{1,\ldots,T-\tau\}$.
We estimate the effective lag-scaling exponent from an unweighted ordinary least-squares (OLS) fit over these six lag values:
\begin{equation}\label{eq:H_regression}
\ln S_{2}^{\mathrm{pool}}(\tau)=2H\ln\tau+\mathrm{const}.
\end{equation}
This fit yields $\hat{H}$.
The dyadic range lies well below the full observation horizon and covers daily to approximately monthly fluctuations, without treating the weekly period as a separately fitted scale.

Pooling over trajectory episodes and admissible ages stabilises the empirical second-order statistic.
It also combines increments with age-dependent fluctuation prefactors and any residual age dependence in their mean increments.
Accordingly, $\hat{H}$ is interpreted as an effective finite-window estimate of the second-order lag-scaling exponent..pdf
Within the fBM reference description, $\hat{H}<1/2$ corresponds to antipersistent increment scaling.
Such second-order lag scaling does not, however, uniquely identify fBM among correlated processes with the same exponent.

The quoted uncertainty of $\hat{H}$ is one half of the ordinary regression-slope standard error in Eq.~\eqref{eq:H_regression}.
Because only six lag values enter this regression, and because the pooled increments are temporally and cross-sectionally dependent, this uncertainty characterises the log--log fit conditional on the aggregated variance estimates.
It is not a cluster-robust sampling uncertainty for independent trajectories.
The close agreement of the point estimates across the six separately assembled yearly cohorts therefore provides the more relevant diagnostic of empirical reproducibility.

\begin{figure*}[t!]
\centering
\safegraphic[width=0.7\linewidth]{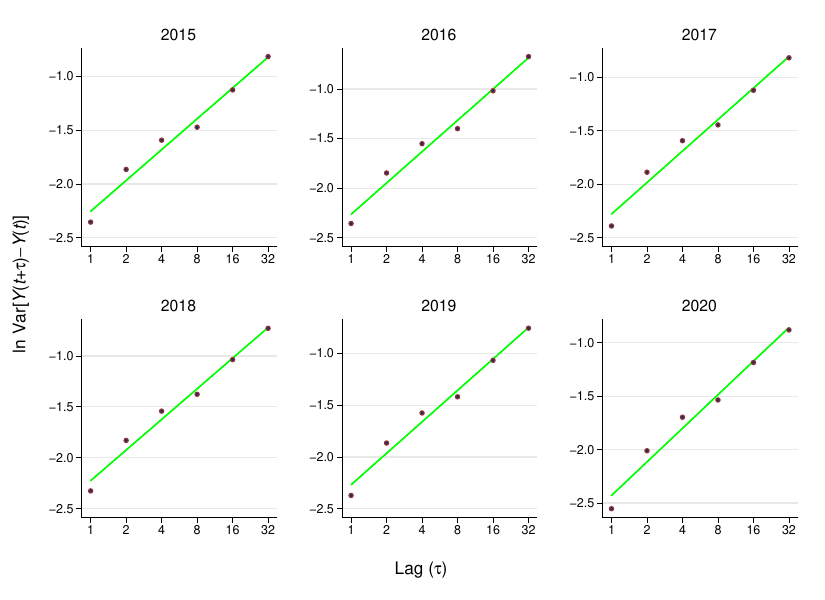}
\caption{Effective lag-scaling exponents across yearly cohorts.
Pooled second-order increment variances
$S_{2}^{\mathrm{pool}}(\tau)$ for dyadic lags
$\tau\in\{1,2,4,8,16,32\}$, calculated separately for each yearly cohort by pooling over trajectory episodes and admissible elapsed days.
The log--log slopes fitted according to Eq.~\eqref{eq:H_regression} yield $2\hat{H}$, with
$\hat{H}\approx0.21$--$0.23$ across the six cohorts.
Within the fBM reference description, these values correspond to antipersistent increment scaling and graph dimensions
$\hat{D}=2-\hat{H}\approx1.77$--$1.79$.
They are effective finite-window estimates because the pooled statistic combines increments with age-dependent fluctuation prefactors.}
\label{fig:est_H_aggr}
\end{figure*}

\subsection{Fitting-range selection and logarithmic variance growth}\label{SI:tc_fit}
For each yearly cohort, let
\begin{equation}
S(t)\coloneq\Var_{k}[Y_{k,t}]
\end{equation}
denote the cross-sectional variance across trajectory episodes at fixed elapsed day $t$.
The logarithmic specification fitted in the Letter is
\begin{equation}\label{eq:SM_log_variance}
S(t)=\gamma\ln t+\mathrm{const}.
\end{equation}
The reported fitting window is determined by a single-breakpoint least-squares procedure applied to $S(t)$ as a function of $\ln t$.
For each candidate split point $s$, two independent OLS lines are fitted on the intervals
$t\in\{1,\ldots,s\}$ and
$t\in\{s+1,\ldots,T\}$.
The split point is constrained by
\begin{equation}
n\leq s\leq T-n, \quad n\coloneq\max\left\{5,\left\lfloor\frac{T}{6}\right\rfloor\right\},
\end{equation}
so that each segment contains at least $n$ observations.
Let $\mathrm{RSS}_{[u,v]}$ denote the residual sum of squares of the OLS line fitted over $t\in\{u,\ldots,v\}$.
The selected endpoint is
\begin{equation}
\tc=\underset{s}{\operatorname{argmin}}\left\{\mathrm{RSS}_{[1,s]}+\mathrm{RSS}_{[s+1,T]}\right\}.
\end{equation}
For $T=365$, $n=60$ and the admissible candidate range is
$60\leq s\leq305$.
The endpoints selected for the six yearly cohorts range from
$\tc=61$ to $\tc=276$, so every reported fitting window spans more than an order of magnitude in elapsed time.

The breakpoint is used only to define the reported finite-window scaling interval $1\leq t\leq\tc$.
The same year-specific value of $\tc$ is then held fixed when estimating the logarithmic variance coefficient $\gamma$, the mean-relaxation coefficient $a$, and the competing variance specification described in Sec.~\ref{SI:compare_powerlaw}.
Consequently, all comparisons within a yearly cohort use identical elapsed days.
The portions of the empirical curves beyond $\tc$ remain displayed in the corresponding figures but are excluded from the subsequent scaling fits and are not assigned to an additional dynamical regime.
Accordingly, $\tc$ is an operational fitting endpoint rather than a physical transition time or an independently identified characteristic timescale.
Its cohort-to-cohort variation is treated as variation in the data-adaptive fitting range, not as evidence for variation in a physical timescale.

Because $\tc$ is selected from the same empirical variance curve subsequently used for fitting, the resulting regression uncertainties and model-comparison statistics are conditional on the selected window.
They do not incorporate uncertainty associated with breakpoint selection and should therefore be interpreted descriptively.
The complementary diagnostic is the absence, over the selected interval, of the systematic convexity or flattening predicted away from marginality (Sec.~\ref{SI:log_slope}).

\subsection{Comparison with a power-law variance fit}\label{SI:compare_powerlaw}
We compare the logarithmic specification in Eq.~\eqref{eq:SM_log_variance} with the two-parameter power law
\begin{equation}\label{eq:SM_power_variance}
S(t)=c_{1}t^{c_{2}}.
\end{equation}
For each yearly cohort, both models are fitted by least squares to the untransformed variance curve $S(t)$ over the identical interval
$1\leq t\leq\tc$.
Thus the residual sums of squares for the two specifications are evaluated on the same response scale and from the same number of observations.

For a descriptive comparison under the usual Gaussian least-squares likelihood, a model with residual sum of squares $\mathrm{RSS}$ is assigned the Akaike information criterion
\begin{equation}
\mathrm{AIC}=n_{\mathrm{fit}}\ln\left(\frac{\mathrm{RSS}}{n_{\mathrm{fit}}}\right)+2k.
\end{equation}
Here $n_{\mathrm{fit}}=\tc$ and $k=2$ for both specifications.
Terms common to both models have been omitted.
Both specifications have two fitted parameters, so their AIC difference is determined by their residual sums of squares on the common fitting window.

As shown in Table~\ref{tab:model_comparison}, the logarithmic specification gives the lower RSS and AIC in every yearly cohort.
The comparison does not establish that no small positive power can approximate a logarithm over a sufficiently restricted interval.
Rather, it shows that the two-parameter positive-power specification in Eq.~\eqref{eq:SM_power_variance} is systematically disfavoured relative to logarithmic growth on the data-adaptive windows used in the Letter.
This conditional comparison is supplemented by the cross-year reproducibility of the logarithmic slopes and by the curvature diagnostic in Sec.~\ref{SI:log_slope}.

\begin{table}[t!]
\centering
\caption{Descriptive comparison of logarithmic and power-law variance fits.
Both specifications are fitted to the untransformed cross-sectional variance $S(t)$ over the same data-adaptive interval $1\leq t\leq\tc$.
Here $\Delta\mathrm{AIC}\coloneq\mathrm{AIC}_{\mathrm{pow}}-\mathrm{AIC}_{\mathrm{log}}$, so positive values favour the logarithmic specification.
The logarithmic model yields the lower RSS and AIC in every yearly cohort.
The statistics are conditional on the breakpoint selected from the same variance curve.}
\label{tab:model_comparison}
\setlength{\tabcolsep}{5.5pt}
\vspace{1em}
\begin{ruledtabular}
\begin{tabular}{c@{\hspace{1.7em}}rr@{\hspace{1.7em}}rr@{\hspace{1.7em}}r}
\raisebox{0.2em}{Year}
& \raisebox{0.2em}{RSS$_{\mathrm{log}}$}
& \raisebox{0.2em}{RSS$_{\mathrm{pow}}$}
& \raisebox{0.2em}{AIC$_{\mathrm{log}}$}
& \raisebox{0.2em}{AIC$_{\mathrm{pow}}$}
& \raisebox{0.2em}{$\Delta$AIC} \\
\colrule
2015 & 0.0153 & 0.3676 & $-700.1$  & $-439.4$  & 260.7 \\
2016 & 0.1095 & 1.2628 & $-2157.6$ & $-1482.8$ & 674.8 \\
2017 & 0.0325 & 0.3388 & $-455.8$  & $-312.8$  & 143.0 \\
2018 & 0.0282 & 0.3548 & $-464.3$  & $-310.0$  & 154.4 \\
2019 & 0.0790 & 0.9978 & $-1179.8$ & $-784.1$  & 395.7 \\
2020 & 0.0149 & 0.3605 & $-741.0$  & $-466.8$  & 274.2 \\
\end{tabular}
\end{ruledtabular}
\end{table}

\subsection{Cross-year empirical scaling estimates}\label{SI:cross_year_estimates}
\begin{figure*}[t!]
\centering
\safegraphic[width=0.7\linewidth]{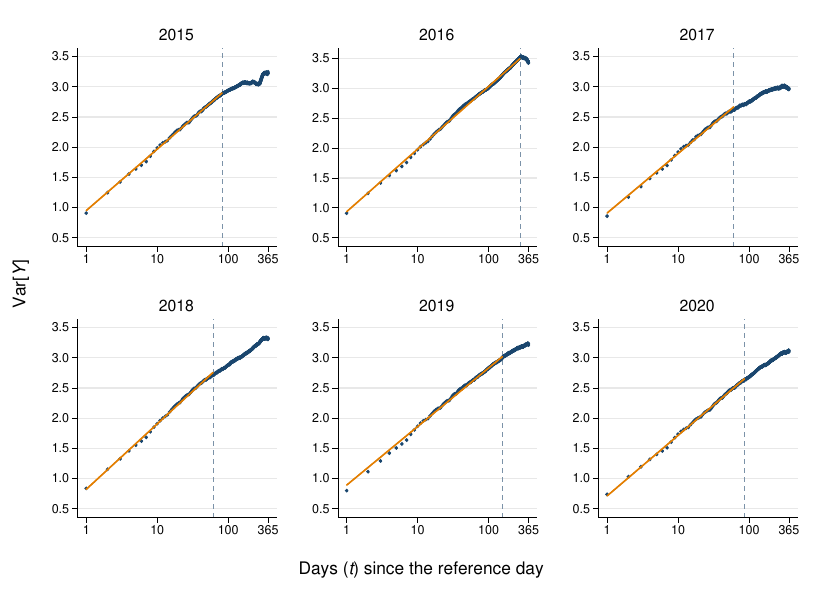}
\caption{Logarithmic growth of ensemble dispersion across yearly cohorts.
The cross-sectional variance $S(t)=\Var_{k}[Y_{k,t}]$ is plotted against $\ln t$ separately for each reference-year cohort.
The solid lines show the fits $S(t)=\gamma\ln t+\mathrm{const}$ over the year-specific intervals $1\leq t\leq\tc$.
The dotted vertical lines mark the fitted endpoints $t=\tc$.
Within the selected windows, the approximately constant slopes and the absence of systematic convexity or flattening are consistent with marginal logarithmic spreading in the scale-free model developed in the Letter.}
\label{fig:est_gamma_aggr}
\end{figure*}

The variance-growth coefficient $\gamma$ is estimated by OLS using Eq.~\eqref{eq:SM_log_variance} over the year-specific interval $1\leq t\leq\tc$.
The endpoint $\tc$ is selected as described in Sec.~\ref{SI:tc_fit}.
Figure~\ref{fig:est_gamma_aggr} shows the resulting variance curves and fits for all six yearly cohorts.
The estimates occupy the narrow range
$\hat{\gamma}=0.424$--$0.468$ despite substantial differences in cohort size and in the selected fitting endpoint.

\begin{figure*}[t!]
\centering
\safegraphic[width=0.7\linewidth]{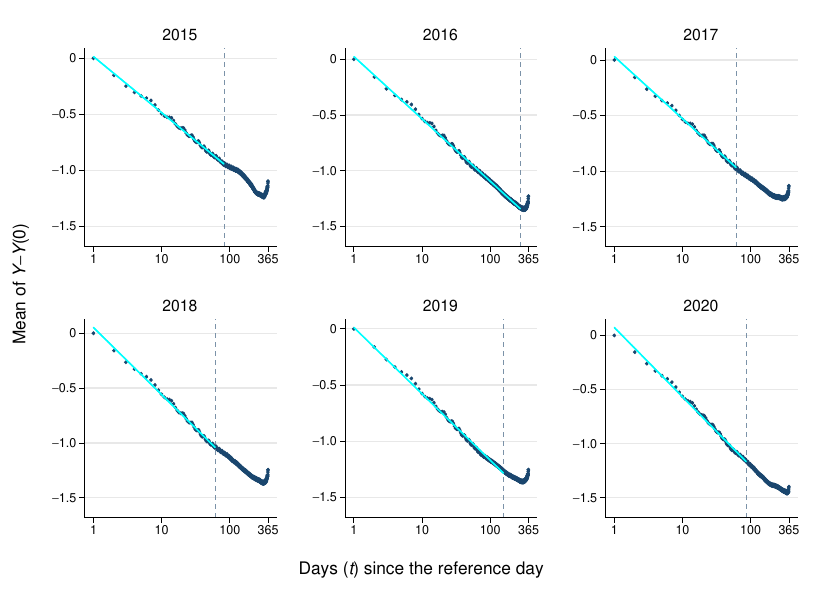}
\caption{Logarithmic relaxation of mean attention across yearly cohorts.
The ensemble mean $M(t)=\left\langle Y_{k,t}-Y_{k,0}\right\rangle_{k}$ is plotted against $\ln t$ separately for each reference-year cohort.
The solid lines show the fits $M(t)=-a\ln t+\mathrm{const}$ over the same year-specific intervals $1\leq t\leq\tc$ used for the variance analysis.
The dotted vertical lines mark $t=\tc$.
The approximately linear dependence on $\ln t$ corresponds to an effective cohort-level relaxation rate proportional to $-1/t$.}
\label{fig:est_a_aggr}
\end{figure*}

For the first ensemble moment, define
\begin{equation}
M(t)\coloneq\left\langle Y_{k,t}-Y_{k,0}\right\rangle_{k},
\end{equation}
where $Y_{k,0}$ is the selection-day reference value for trajectory episode $k$.
The cohort-level decay coefficient $a$ is estimated over the same interval as
\begin{equation}
M(t)=-a\ln t+\mathrm{const},\quad 1\leq t\leq\tc.
\end{equation}
The additive constant allows for the difference between the selection-day
reference and the first post-selection observation.
Under this sign convention, $a>0$ denotes declining mean attention.
The corresponding yearly curves are shown in Fig.~\ref{fig:est_a_aggr}.

The three reported coefficients describe distinct empirical properties.
The effective Hurst exponent $H$ characterises the finite-window lag scaling of temporal increments, $\gamma$ measures the logarithmic growth rate of the cross-sectional variance, and $a$ measures logarithmic relaxation of the distribution centre.
Thus $H$, $\gamma$, and $a$ are not alternative estimates of a common parameter.
As summarised in Table~\ref{tab:summary}, $\hat{H}$ and $\hat{\gamma}$ remain comparatively stable across yearly cohorts, whereas $\hat{a}$ shows a modest overall upward tendency despite nonmonotonic year-to-year variation.
This tendency corresponds to faster cohort-level mean relaxation in the later years.

\begin{table*}[t]
\centering
\caption{Empirical scaling estimates by reference-year cohort.
The table reports the fitted endpoint $\tc$, the effective Hurst exponent $\hat{H}$ obtained from pooled increment scaling, the logarithmic variance-growth coefficient $\hat{\gamma}$, and the cohort-level decay coefficient $\hat{a}$.
Quoted uncertainties are ordinary regression-fit standard errors conditional on the corresponding aggregated curves and, for $\hat{\gamma}$ and $\hat{a}$, on the selected endpoint $\tc$.
They do not incorporate breakpoint-selection uncertainty or temporal and cross-episode dependence and should not be interpreted as cluster-robust sampling uncertainties for independent articles.}
\label{tab:summary}
\setlength{\tabcolsep}{8pt}
\vspace{1em}
\begin{ruledtabular}
\begin{tabular}{crccc}
\raisebox{0.2em}{Reference year}
& \raisebox{0.2em}{$\tc$}
& \raisebox{0.2em}{$\hat{H}$}
& \raisebox{0.2em}{$\hat{\gamma}$}
& \raisebox{0.2em}{$\hat{a}$} \\
\colrule
2015 & $82$  & $0.207\pm0.016$ & $0.444\pm0.002$ & $0.216\pm0.002$ \\
2016 & $276$ & $0.228\pm0.016$ & $0.456\pm0.001$ & $0.242\pm0.001$ \\
2017 & $61$  & $0.212\pm0.016$ & $0.430\pm0.003$ & $0.241\pm0.003$ \\
2018 & $61$  & $0.217\pm0.015$ & $0.468\pm0.003$ & $0.264\pm0.003$ \\
2019 & $156$ & $0.219\pm0.015$ & $0.424\pm0.002$ & $0.255\pm0.002$ \\
2020 & $86$  & $0.227\pm0.017$ & $0.434\pm0.002$ & $0.277\pm0.002$ \\
\end{tabular}
\end{ruledtabular}
\end{table*}

\section{Theory and Diagnostics}\label{SI:theory}
This section defines the canonical Wiener integral with respect to fBM used in the Letter, derives the three variance regimes, and relates the local increment scaling and logarithmic-slope diagnostics to the empirical estimators introduced above.

\subsection{Wiener integration with fractional Brownian motion}\label{SI:stochastic_calculus}
Let $B_{H}(t)$ be a centred fBM with
$B_{H}(0)=0$, Hurst exponent $H\in(0,1)$, and covariance
\begin{equation}\label{eq:fBM_covariance}
R_{H}(s,u)\coloneq\mathbb{E}\!\left[B_{H}(s)B_{H}(u)\right]=\frac{1}{2}\bigl(s^{2H}+u^{2H}-|s-u|^{2H}\bigr).
\end{equation}
For $H\neq1/2$, fBM is not a semimartingale, so the stochastic integral in the model is not defined as an It\^{o} integral.
Since the fluctuation amplitude is deterministic, we define the stochastic term as the canonical Wiener integral with respect to $B_{H}$ \cite{Nualart06}.
For interval indicators, define the inner product
\begin{align}
&\left\langle\mathbf{1}_{(a,b]},\mathbf{1}_{(c,d]}\right\rangle_{\mathcal{H}_{H}}\notag\\
&\hspace{1em}\coloneq\mathbb{E}\Bigl[\bigl(B_{H}(b)-B_{H}(a)\bigr)\bigl(B_{H}(d)-B_{H}(c)\bigr)\Bigr].
\end{align}
The completion of the linear span of these indicators under this inner product defines the canonical Hilbert space $\mathcal{H}_{H}$ associated with $B_{H}$.
For a deterministic step function
\begin{equation}
\varphi(s)=\sum_{j=1}^{n}c_{j}\mathbf{1}_{(u_{j},v_{j}]}(s),
\end{equation}
the Wiener integral is defined by
\begin{equation}
\int\varphi(s)\dd{B_{H}(s)}\coloneq\sum_{j=1}^{n}c_{j}\left[B_{H}(v_{j})-B_{H}(u_{j})\right].
\end{equation}
This map extends by mean-square completion to every deterministic integrand
$\varphi\in\mathcal{H}_{H}$ and satisfies the isometry
\begin{equation}\label{eq:Wiener_isometry}
\mathbb{E}\left[\left(\int_{t_{0}}^{t}\varphi(s)\dd{B_{H}(s)}\right)^{2}\right]=\left\|\varphi\mathbf{1}_{(t_{0},t]}\right\|_{\mathcal{H}_{H}}^{2}.
\end{equation}
Because the integral is centred, the left-hand side of Eq.~\eqref{eq:Wiener_isometry} is also its variance.
The deterministic power-law amplitudes used in the Letter are smooth and bounded on every finite interval $[t_{0},t]$ with $t_{0}>0$ and are admissible integrands in this construction.
Consequently, the stochastic accumulations $I(t;t_{0})$ and $I_{\rho}(t)$ used in the Letter are well-defined centred Gaussian variables.
Their variances and covariances between logarithmic time intervals are determined by the fBM covariance through this Hilbert-space construction, providing the basis for the scaling analysis in Sec.~\ref{SI:scaling_log}.

The canonical Hilbert-space definition applies for $H<1/2$ as well as for $H>1/2$.
In particular, it does not require the formal derivative covariance of fBM to be treated as an ordinary integrable kernel, which would be inappropriate in the antipersistent range $H<1/2$ relevant to the empirical estimates.

For completeness, we restate Eq.~\eqref{eq:log_model} of the Letter in level form:
\begin{equation}\label{eq:SM_log_integral_model}
Y(t)=Y(t_{0})+\int_{t_{0}}^{t}\alpha(s)\dd{s}+\int_{t_{0}}^{t}\beta(s)\dd{B_{H}(s)},
\end{equation}
where $\alpha(t)$ and $\beta(t)$ are deterministic functions.
The positive activity process is subsequently defined by
\begin{equation}
X(t)\coloneq\exp Y(t).
\end{equation}
No It\^{o}, Stratonovich, or other stochastic chain rule is invoked in passing between $Y$ and $X$.
The positive reference time $t_{0}$ avoids the singular coefficients that would arise at the origin and changes finite-time constants without affecting the asymptotic scaling classification.
The fBM driver is used here as a parsimonious Gaussian reference process for the measured second-order scaling, not as a uniquely identified microscopic mechanism.

Corresponding to Eq.~\eqref{eq:I_def} of the Letter, the stochastic accumulation is
\begin{equation}\label{eq:SM_stochastic_accumulation}
I(t;t_{0})\coloneq\int_{t_{0}}^{t}\beta(s)\dd{B_{H}(s)}.
\end{equation}
Because the drift in Eq.~\eqref{eq:SM_log_integral_model} is deterministic,
\begin{equation}
\Var[Y(t)-Y(t_{0})]=\Var[I(t;t_{0})].
\end{equation}
The variance of the level satisfies the more general relation
\begin{align}\label{eq:SM_level_variance_relation}
\Var[Y(t)]&=\Var[Y(t_{0})]+\Var[I(t;t_{0})]\notag\\
&{\hspace{1em}}+2\Cov[Y(t_{0}),I(t;t_{0})].
\end{align}
Under the reduced ensemble assumption that the initial condition is uncorrelated with the subsequent stochastic accumulation, the covariance term vanishes and the initial cross-sectional variance contributes only an additive baseline.

\subsection{Logarithmic-time derivation of the variance regimes}\label{SI:scaling_log}
For the scale-free amplitude in Eq.~\eqref{eq:beta_power} of the Letter, the stochastic accumulation in Eq.~\eqref{eq:SM_stochastic_accumulation} becomes
\begin{equation}
I(t;t_{0})=\frac{b}{t_{0}^{H}}\int_{t_{0}}^{t}\left(\frac{s}{t_{0}}\right)^{-\eta}\dd{B_{H}(s)}.
\end{equation}
The balance between self-affine accumulation and ageing attenuation is measured by
$\xi\coloneq H-\eta$, as introduced in Eq.~\eqref{eq:xi} of the Letter.
The deterministic drift affects the ensemble mean but not $\Var[I(t;t_{0})]$.
Its relation to the level variance $\Var[Y(t)]$ is given by Eq.~\eqref{eq:SM_level_variance_relation}.
The additional variance and covariance terms generated by a latent trajectory-specific drift are examined separately in Sec.~\ref{SI:gmm_scope}.

Choose an arbitrary fixed ratio $\rho>1$ and introduce logarithmically spaced times
\begin{equation}
t_{n}=t_{0}\rho^{n},\quad n=0,1,2,\ldots.
\end{equation}
The stochastic contribution accumulated over the $n$th logarithmic interval is
\begin{equation}\label{eq:SM_Zn}
Z_{n}\coloneq I(t_{n+1};t_{0})-I(t_{n};t_{0})=\frac{b}{t_{0}^{H}}\int_{t_{n}}^{t_{n+1}}\left(\frac{s}{t_{0}}\right)^{-\eta}\dd{B_{H}(s)}.
\end{equation}
Because $t_{n+1}=\rho t_{n}$, this is precisely $Z_{n}=I_{\rho}(t_{n})$ in the notation of
Eq.~\eqref{eq:log_interval_contr} of the Letter.
By the $H$-self-similarity of fBM, the processes $\{B_{H}(cs)\}_{s\geq0}$ and $\{c^{H}B_{H}(s)\}_{s\geq0}$ have the same finite-dimensional distributions for every $c>0$.
A rescaling of the integration interval therefore gives
\begin{equation}
Z_{n}\overset{\mathrm{d}}{=}\rho^{n(H-\eta)}Z_{0}=\rho^{n\xi}Z_{0},
\end{equation}
where $\overset{\mathrm{d}}{=}$ denotes equality in distribution.
Define the normalised logarithmic-interval sequence
\begin{equation}
W_{n}\coloneq\rho^{-n\xi}Z_{n}.
\end{equation}
Joint self-similarity of $B_{H}$ together with the homogeneous fluctuation amplitude implies that $\{W_{n}\}$ is a centred stationary Gaussian sequence in the logarithmic index $n$.
Let
\begin{equation}
\Gamma(\ell)\coloneq\Cov(W_{0},W_{\ell}),\quad\ell\geq0.
\end{equation}
Direct evaluation using the fBM covariance in Eq.~\eqref{eq:fBM_covariance} shows that, for separated logarithmic intervals,
\begin{equation}
\Gamma(\ell)=O\bigl(\rho^{-(1-H)\ell}\bigr),\quad\ell\to\infty.
\end{equation}
For $H=1/2$, the corresponding Wiener integrals over nonoverlapping Brownian intervals are independent, so the off-diagonal covariances vanish.
Thus, for every $H\in(0,1)$, the covariance sequence is absolutely summable.

At the marginal balance $\eta=H$, the long-run variance of the logarithmic-interval sequence is
\begin{equation}
\Lambda_{H,\rho}\coloneq\Gamma(0)+2\sum_{\ell=1}^{\infty}\Gamma(\ell).
\end{equation}
For the nondegenerate fBM block integral in Eq.~\eqref{eq:SM_Zn}, direct evaluation of the covariance gives
\begin{equation}\label{eq:SM_positive_long_run_variance}
0<\Lambda_{H,\rho}<\infty.
\end{equation}
The strict positivity in Eq.~\eqref{eq:SM_positive_long_run_variance}, rather than covariance summability alone, prevents complete cancellation of successive logarithmic-interval contributions.

At $t=t_{N}$, the stochastic integral is the weighted partial sum
\begin{equation}\label{eq:SM_sum_intervals}
I(t_{N};t_{0})=\sum_{n=0}^{N-1}Z_{n}=\sum_{n=0}^{N-1}\rho^{n\xi}W_{n}.
\end{equation}
The asymptotic behaviour of its variance depends on the sign of $\xi$.
For the three cases below, denote this variance by
\begin{equation}
V_{I}(t)\coloneq\Var[I(t;t_{0})].
\end{equation}

\SIrunin{Amplifying regime ($\xi>0$).}
The weights increase geometrically, so the most recent logarithmic intervals dominate the sum.
Stationarity of $\{W_{n}\}$ and the geometric decay of its covariance give
\begin{equation}\label{eq:SM_supercritical}
V_{I}(t_{N})\sim C_{+}\rho^{2N\xi}=C_{+}\left(\frac{t_{N}}{t_{0}}\right)^{2\xi},\quad C_{+}>0.
\end{equation}
Thus the stochastic contribution to the ensemble variance grows algebraically with age.

\SIrunin{Saturating regime ($\xi<0$).}
The geometrically weighted series in Eq.~\eqref{eq:SM_sum_intervals} converges in mean square.
Consequently,
\begin{equation}\label{eq:SM_subcritical}
V_{I}(t_{N})\to C_{-},\quad 0<C_{-}<\infty.
\end{equation}
Its variance therefore approaches a finite asymptotic limit.

\SIrunin{Marginal regime ($\xi=0$).}
At the balance point, all logarithmic intervals enter with equal weight:
\begin{equation}
I(t_{N};t_{0})=\sum_{n=0}^{N-1}W_{n}.
\end{equation}
Its variance is
\begin{align}
V_{I}(t_{N})&=N\Gamma(0)+2\sum_{\ell=1}^{N-1}(N-\ell)\Gamma(\ell)\notag\\
&=\Lambda_{H,\rho}N+K_{H,\rho}+O\bigl(\rho^{-(1-H)N}\bigr),
\label{eq:SM_marginal_N}
\end{align}
as $N\to\infty$.
Here
\begin{equation}
K_{H,\rho}\coloneq-2\sum_{\ell=1}^{\infty}\ell\Gamma(\ell)
\end{equation}
is finite because the covariance sequence decays geometrically.
The logarithmic index $N$ is related to physical time by
\begin{equation}
N=\frac{\ln(t_{N}/t_{0})}{\ln\rho}.
\end{equation}
Substitution into Eq.~\eqref{eq:SM_marginal_N} gives
\begin{equation}\label{eq:SM_marginal_log}
V_{I}(t_{N})=C_{0}\ln\left(\frac{t_{N}}{t_{0}}\right)+K_{H,\rho}+O\!\left[\left(\frac{t_{N}}{t_{0}}\right)^{-(1-H)}\right],
\end{equation}
where
\begin{equation}
C_{0}\coloneq\frac{\Lambda_{H,\rho}}{\ln\rho}>0.
\end{equation}
Because $Z_{n}$ and $W_{n}$ depend linearly on $b$, one has $\Lambda_{H,\rho}\propto b^{2}$ and hence $C_{0}\propto b^{2}$.
The ratio $\rho$ only determines how the same stochastic integral is partitioned into logarithmic intervals, and $V_{I}(t)$ itself does not depend on this choice.
Consequently, although $\Lambda_{H,\rho}$ depends on $\rho$, the combination $C_{0}=\Lambda_{H,\rho}/\ln\rho$ does not, and it gives the coefficient of $\ln(t/t_{0})$ in the asymptotic physical-time variance.

The three asymptotic regimes can therefore be summarised as
\begin{equation}
V_{I}(t)\sim
\begin{cases}
~C_{+}(t/t_{0})^{2\xi}, & \xi>0,\\[3pt]
~C_{0}\ln(t/t_{0}), & \xi=0,\\[3pt]
~C_{-}, & \xi<0.
\end{cases}
\end{equation}
The marginal logarithm therefore arises from comparable-scale fluctuations on successive logarithmic time intervals whose covariance sequence is summable and whose long-run variance is positive.
It separates algebraic broadening from saturation at a finite variance.

\subsection{Logarithmic-slope signatures}\label{SI:log_slope}
For a fixed scale ratio $\rho>1$, define the effective slope over one logarithmic interval by
\begin{equation}\label{eq:SM_gamma_rho}
\gamma_{\rho}(t)\coloneq\frac{V_{I}(\rho t)-V_{I}(t)}{\ln\rho}.
\end{equation}
This finite-scale quantity directly measures the change in variance per logarithmic time interval.
When $V_{I}(t)$ is differentiable, $\gamma_{\rho}(t)$ approaches
$\dd{V_{I}(t)}/\dd{\ln t}$ as $\rho\to 1$.
The finite-scale definition in Eq.~\eqref{eq:SM_gamma_rho} is sufficient to distinguish the three asymptotic regimes without requiring differentiability of the variance curve.

\SIrunin{Amplifying regime ($\xi>0$).}
Using Eq.~\eqref{eq:SM_supercritical} in Eq.~\eqref{eq:SM_gamma_rho} gives
\begin{equation}
\gamma_{\rho}(t)\sim C_{+}\frac{\rho^{2\xi}-1}{\ln\rho}\left(\frac{t}{t_{0}}\right)^{2\xi}.
\end{equation}
The effective logarithmic slope therefore increases algebraically with age.
Equivalently, the variance curve is asymptotically convex when plotted against $\ln(t/t_{0})$.
This behaviour defines the liquid-like amplification side of the dynamical analogy used in the Letter.

\SIrunin{Saturating regime ($\xi<0$).}
Equation~\eqref{eq:SM_subcritical} implies
\begin{equation}\label{eq:SM_gamma_subcritical}
\gamma_{\rho}(t)\to 0\quad(t\to\infty).
\end{equation}
Thus the change in variance over each successive logarithmic interval vanishes asymptotically.
When the finite limiting variance is approached from below, the slope is positive and decreases towards zero; more generally, Eq.~\eqref{eq:SM_gamma_subcritical} establishes asymptotic flattening without requiring a particular power law for the approach to the limit.
This behaviour defines the frozen-saturation side of the analogy.

\SIrunin{Marginal regime ($\xi=0$).}
Equation~\eqref{eq:SM_marginal_log} gives
\begin{equation}
\gamma_{\rho}(t)\to C_{0}>0\quad(t\to\infty).
\end{equation}
The variance therefore increases by the same asymptotic amount over successive logarithmic time intervals.
The variance curve approaches a straight line with nonzero slope when plotted against $\ln(t/t_{0})$, rather than becoming progressively steeper or flatter.

\begin{table*}[t!]
\centering
\caption{Asymptotic logarithmic-slope signatures of the three dynamical regimes.
The effective slope $\gamma_{\rho}(t)$ distinguishes algebraic amplification, marginal logarithmic spreading, and saturation through the corresponding shape of the variance curve against $\ln(t/t_{0})$.}
\label{tab:SM_slope_signatures}
\setlength{\tabcolsep}{10pt}
\vspace{0.5em}
\begin{ruledtabular}
\begin{tabular}{@{}lll@{}}
\raisebox{0.2em}{Regime} & \raisebox{0.2em}{$\gamma_{\rho}(t)$} & \raisebox{0.2em}{Shape versus $\ln(t/t_{0})$} \\
\colrule
$\xi>0$~~(amplification) & increases algebraically & asymptotically convex \\
$\xi=0$~~(marginal) & approaches $C_{0}>0$ & asymptotically linear \\
$\xi<0$~~(saturation) & approaches zero & asymptotically flat \\
\end{tabular}
\end{ruledtabular}
\end{table*}

Table~\ref{tab:SM_slope_signatures} summarises these asymptotic predictions.
Over a finite observation window, the corresponding curve shapes provide a model-based consistency check rather than an independent test of marginality.
An increasing slope indicates amplification, a slope approaching zero indicates saturation, and a positive asymptotically constant slope indicates marginal spreading.

\subsection{Increment scaling under age-dependent amplitude}\label{SI:scaling_H}
Consider the stochastic component
\begin{equation}\label{eq:SM_increment}
\Delta_{\tau}I(t)\coloneq I(t+\tau;t_{0})-I(t;t_{0})
=\frac{b}{t_{0}^{H}}\int_{t}^{t+\tau}\left(\frac{s}{t_{0}}\right)^{-\eta}\dd{B_{H}(s)}.
\end{equation}
For a local increment satisfying $0<\tau\ll t$, the deterministic amplitude varies slowly over the interval $[t,t+\tau]$.
Uniformly over this interval, $(s/t_{0})^{-\eta}=(t/t_{0})^{-\eta}[1+O(\tau/t)]$.
The Wiener integral in Eq.~\eqref{eq:SM_increment} can therefore be written in mean square as
\begin{equation}\label{eq:SM_increment_approx}
\Delta_{\tau}I(t)=\frac{b}{t_{0}^{H}}\left(\frac{t}{t_{0}}\right)^{-\eta}\left[B_{H}(t+\tau)-B_{H}(t)\right]+R(t,\tau),
\end{equation}
where the remainder satisfies
\begin{equation}
\left\|R(t,\tau)\right\|_{2}=O\left(\frac{\tau}{t}\right)\frac{b}{t_{0}^{H}}\left(\frac{t}{t_{0}}\right)^{-\eta}\tau^{H}.
\end{equation}
By Eq.~\eqref{eq:S_H_scaling}, the fBM increment in Eq.~\eqref{eq:SM_increment_approx} has variance $\tau^{2H}$.
Consequently,
\begin{equation}\label{eq:SM_increment_variance}
\Var[\Delta_{\tau}I(t)]=\frac{b^{2}}{t_{0}^{2H}}\left(\frac{t}{t_{0}}\right)^{-2\eta}\tau^{2H}\left[1+O\left(\frac{\tau}{t}\right)\right].
\end{equation}
Thus the local lag exponent is $2H$, whereas the age-dependent amplitude changes the leading prefactor through $t^{-2\eta}$.

The deterministic drift contributes
\begin{equation}
\Delta_{\tau}A(t)\coloneq\int_{t}^{t+\tau}\alpha(s)\dd{s}.
\end{equation}
At fixed $t$, this deterministic contribution changes the mean increment but does not contribute to its variance across realisations of the reduced stochastic model.
For the scale-free drift $\alpha(t)=-a/t$ used in the Letter,
\begin{equation}
\Delta_{\tau}A(t)=-a\ln\left(1+\frac{\tau}{t}\right)=-a\frac{\tau}{t}+O\left(\frac{\tau^{2}}{t^{2}}\right).
\end{equation}
The drift contribution is therefore locally smooth.
At the marginal balance $\eta=H$, the ratio of its magnitude to the root-mean-square stochastic increment is $(a/b)(\tau/t)^{1-H}[1+O(\tau/t)]$ and hence vanishes as $\tau/t\to 0$.

Equation~\eqref{eq:SM_increment_variance} is a local fixed-age relation.
The empirical statistic $S_{2}^{\mathrm{pool}}(\tau)$ in Eq.~\eqref{eq:SM_pooled_increment_variance}, by contrast, combines increments over all admissible ages and trajectory episodes.
It therefore mixes age-dependent fluctuation prefactors and incorporates any residual age variation in the mean increments, rather than averaging fixed-age conditional variances alone.
Accordingly, it should not be identified exactly with the fixed-age variance in Eq.~\eqref{eq:SM_increment_variance}.
The fitted $\hat{H}$ is interpreted as an effective finite-window estimate of the underlying second-order lag exponent.
Its narrow variation across the six yearly cohorts provides the empirical reproducibility check used in the Letter.

Within the fBM reference description, the estimates $\hat{H}<1/2$ correspond to antipersistent increment scaling.
The empirical relation identifies the second-order lag exponent but does not uniquely distinguish fBM from other correlated processes with the same local scaling.

\subsection{Geometric and correlation interpretation of the Hurst exponent}\label{SI:geometric}
For fBM with $H\in(0,1)$, the graph of a sample path has almost-sure fractal dimension
\begin{equation}
D=2-H
\end{equation}
under the standard Euclidean graph metric \cite{Falconer14}.
Within this reference description, the yearly estimates $\hat{H}=0.207$--$0.228$ correspond to
\begin{equation}
\hat{D}=2-\hat{H}=1.772\text{--}1.793.
\end{equation}
At the precision warranted by the regression uncertainties, this corresponds to $\hat{D}\approx1.77$--$1.79$.
For comparison, Brownian motion has $H=1/2$ and $D=3/2$.
The empirical estimates $\hat{H}<1/2$ therefore correspond, within the fBM reference description, to graphs rougher than Brownian paths.
This correspondence is an interpretation of the fitted second-order exponent and is not an independent fractal-dimension estimate from the finite daily trajectories.

The same range $H<1/2$ also implies negatively correlated successive fBM increments.
For two adjacent increments of equal duration $\tau$, write
\begin{equation}
\Delta_{\tau}B_{H}(t)\coloneq B_{H}(t+\tau)-B_{H}(t).
\end{equation}
Using Eq.~\eqref{eq:fBM_covariance}, their covariance is
\begin{equation}
\Cov[\Delta_{\tau}B_{H}(t),\Delta_{\tau}B_{H}(t+\tau)]=\bigl(2^{2H-1}-1\bigr)\tau^{2H}.
\end{equation}
Because $2^{2H-1}-1<0$ for $H<1/2$, a positive increment is statistically associated with a compensating subsequent increment rather than reinforcement.
This is the antipersistent correlation structure referred to in the Letter.

Within the fBM reference model, $H$ characterises local lag scaling, path roughness, and increment correlations, whereas $\xi=H-\eta$ governs cumulative ensemble dispersion.
Rough trajectories and marginal logarithmic spreading are therefore complementary rather than equivalent signatures of the dynamics.

\section{Trajectory Heterogeneity}\label{SI:analysis}
The ensemble-moment analysis in the Letter establishes reproducible scaling at the population level.
Here we show that this collective regularity coexists with broad, non-Gaussian variation in finite-trajectory relaxation.
We quantify the empirical distribution of effective relaxation slopes and then delimit what can and cannot be inferred from its Gaussian-mixture approximation.

\subsection{Trajectory-level slopes and Gaussian-mixture approximation}\label{SI:gmm}
The mixture analysis concerns fitted trajectory-level decay slopes rather than cumulative page views or terminal logarithmic attention.
Let $t_{j}$ denote the elapsed-day value associated with trajectory column $j$.
For the 2020 calculation, the elapsed-day values run from $1$ through $365$, and the first post-selection day is used as the regression reference time, $t_{0}=1$.
For each trajectory episode $k$, the signed log-time slope is estimated from
\begin{equation}
(\hat{c}_{k},\hat{s}_{k})=\underset{c,s}{\operatorname{argmin}}\sum_{j:t_{j}>t_{0}}\left[Y_{k,t_{j}}-Y_{k,t_{0}}-c-s\ln\left(\frac{t_{j}}{t_{0}}\right)\right]^{2}.
\end{equation}
Thus the regression uses the 364 observations with elapsed days $t_{j}=2,\ldots,365$ relative to the post-selection baseline $Y_{k,1}$.
The intercept $\hat{c}_{k}$ is retained rather than forcing the fitted curve through zero at $t=t_{0}$.
The signed slope $\hat{s}_{k}$ is negative for a trajectory that declines on average with logarithmic age.
To use the same positive-decay convention as the cohort-level coefficient in the Letter, define $\hat{a}_{k}\coloneq-\hat{s}_{k}$.
Accordingly, $\hat{a}_{k}>0$ denotes declining attention and $\hat{a}_{k}<0$ denotes increasing attention over the fitted post-selection horizon.
The calculation uses the elapsed-day values parsed from the trajectory labels before taking logarithms; raw array positions are not used as substitutes for time.

We approximate the empirical density of the effective decay coefficients by the Gaussian mixture
\begin{equation}
p_{M}(a)=\sum_{m=1}^{M}w_{m}\mathcal{N}(a\!\mid\!\mu_{m},\sigma_{m}^{2}),\quad w_{m}>0,\quad\sum_{m=1}^{M}w_{m}=1.
\end{equation}
Candidate models with $M\in\{2,\ldots,10\}$ are fitted by maximum likelihood with component-specific one-dimensional variances.
Each candidate fit uses 10 initialisations, k-means initialisation, a maximum of 1,000 iterations, covariance regularisation $10^{-6}$, and random seed 0.
For a candidate mixture with maximised log likelihood $\hat{\ell}_{M}$, the Bayesian information criterion is
\begin{equation}
\mathrm{BIC}_{M}=-2\hat{\ell}_{M}+q_{M}\ln N,\quad q_{M}=3M-1,
\end{equation}
where $N$ is the number of fitted trajectory episodes.
The parameter count contains $M$ means, $M$ variances, and $M-1$ independent mixture weights.
Because the trajectory episodes are not assumed to be statistically independent, BIC is used here as a descriptive model-order criterion rather than as a formal consistency result.

\begin{figure}[t!]
\centering
\safegraphic[width=0.85\columnwidth]{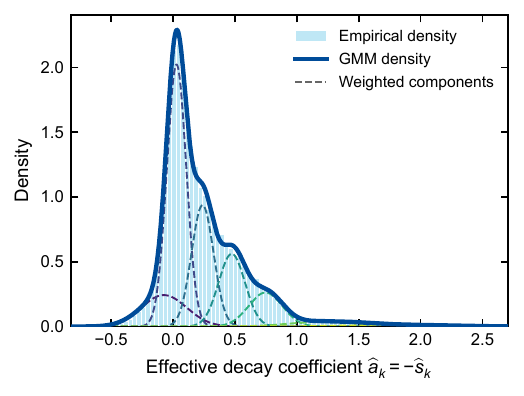}
\caption{Broad heterogeneity of trajectory-level relaxation rates.
Empirical density of the effective decay coefficient
$\hat{a}_{k}=-\hat{s}_{k}$ for the 2020 trajectory episodes, together with the BIC-selected seven-component Gaussian-mixture density and its weighted components.
Values near $\hat{a}_{k}=0$ correspond to slowly relaxing trajectories, whereas larger positive values correspond to more rapid decline.
The mixture provides a smooth description of the empirical density rather than a classification into discrete article types.}
\label{fig:gmm_density_2020}
\end{figure}

\begin{figure}[t!]
\centering
\safegraphic[width=0.83\columnwidth]{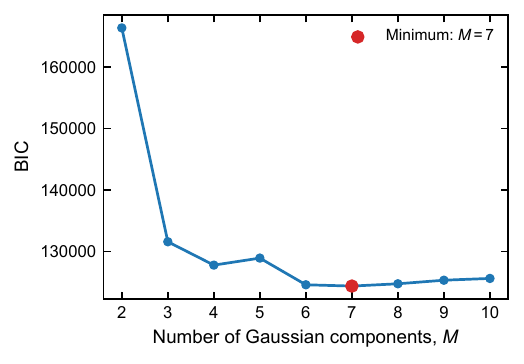}
\caption{Model-order selection for the trajectory-level Gaussian mixture.
BIC for candidate component numbers $M\in\{2,\ldots,10\}$ fitted to the 2020 effective decay coefficients $\{\hat{a}_{k}\}$.
Each candidate model is fitted using 10 initialisations with the common implementation settings specified in Sec.~\ref{SI:gmm}.
The minimum occurs at $M=7$ and selects the descriptive density shown in Fig.~\ref{fig:gmm_density_2020}.
The selected component number describes the flexibility required to approximate the observed density and does not identify discrete latent article classes.}
\label{fig:gmm_bic_2020}
\end{figure}

All $N=365{,}930$ retained 2020 trajectory episodes enter the slope analysis; no additional episode is excluded because all fitted slope values are finite.
Figure~\ref{fig:gmm_density_2020} shows the broad, markedly non-Gaussian empirical density together with its BIC-selected seven-component approximation.
The minimum BIC occurs at $M=7$ for the 2020 cohort, as shown in Fig.~\ref{fig:gmm_bic_2020}.
The empirical distribution spans slowly relaxing, `evergreen' trajectories with $\hat{a}_{k}$ near zero and rapidly fading, breaking-news-like trajectories with substantially larger positive decay coefficients.
Negative values of $\hat{a}_{k}$ correspond to episodes whose fitted attention increases over the observation window, including delayed or recurrent activity.
These descriptions refer to observed trajectory shapes rather than distinct mechanisms or intrinsic article types.
The fitted mixture provides a flexible representation of the empirical density; selection of $M=7$ does not imply the existence of seven discrete article classes.

The mixture-weighted mean is
\begin{equation}\label{eq:SM_gmm_mean}
\overline{a}_{\mathrm{mix}}=\sum_{m=1}^{M}w_{m}\mu_{m}.
\end{equation}
The corresponding mixture variance is
\begin{equation}
\sigma_{\mathrm{mix}}^{2}=\sum_{m=1}^{M}w_{m}\left[\sigma_{m}^{2}+\left(\mu_{m}-\overline{a}_{\mathrm{mix}}\right)^{2}\right].
\end{equation}
The resulting sample and mixture moments are reported in Table~\ref{tab:GMM}.

\begin{table}[t!]
\centering
\caption{Summary of the trajectory-level Gaussian-mixture analysis for 2020.
The effective coefficient follows the positive-decay convention $\hat{a}_{k}=-\hat{s}_{k}$.
The empirical and mixture moments are reported to six decimal places to display the numerical fidelity of the fitted density.
Their agreement is an implementation and density-approximation check, not evidence for discrete latent classes.}
\label{tab:GMM}
\setlength{\tabcolsep}{9pt}
\vspace{1em}
\begin{ruledtabular}
\begin{tabular}{lc}
\raisebox{0.2em}{Quantity} & \raisebox{0.2em}{Value} \\
\colrule
Trajectory episodes & $365{,}930$ \\
Additional exclusions from slope analysis & $0$ \\
Elapsed-day range & $1$--$365$ \\
Regression-day range & $2$--$365$ \\
Candidate component numbers & $2$--$10$ \\
BIC-selected components & $7$ \\
Empirical mean decay ($\overline{a}_{\mathrm{emp}}$) & $0.249119$ \\
Mixture-weighted mean decay ($\overline{a}_{\mathrm{mix}}$) & $0.249119$ \\
Empirical standard deviation ($\sigma_{\mathrm{emp}}$) & $0.368713$ \\
Mixture standard deviation ($\sigma_{\mathrm{mix}}$) & $0.368715$ \\
\end{tabular}
\end{ruledtabular}
\end{table}

\subsection{Relation between trajectory-level slopes and cohort-level relaxation}\label{SI:reproduce_a}
The trajectory-level decay coefficients and the cohort-level decay coefficient probe related first-order behaviour, but they are calculated over different regression windows.
Their numerical values should therefore not be expected to coincide.
To make the relation explicit, let $J$ denote the number of elapsed days included in a given regression and define
\begin{equation}
x_{j}\coloneq\ln\left(\frac{t_{j}}{t_{0}}\right),\quad\overline{x}\coloneq\frac{1}{J}\sum_{j=1}^{J}x_{j}.
\end{equation}
For a common time grid and common regression weights, the signed trajectory-level OLS slope can be written as
\begin{equation}\label{eq:SM_individual_slope_linear}
\hat{s}_{k}=\frac{\sum_{j=1}^{J}(x_{j}-\overline{x})Y_{k,t_{j}}}{\sum_{j=1}^{J}(x_{j}-\overline{x})^{2}}.
\end{equation}
Subtracting $Y_{k,t_{0}}$ or any other trajectory-specific constant does not change Eq.~\eqref{eq:SM_individual_slope_linear}, because $\sum_{j=1}^{J}(x_{j}-\overline{x})=0$.
Let
\begin{equation}
\overline{Y}_{t_{j}}\coloneq \frac{1}{N}\sum_{k=1}^{N}Y_{k,t_{j}}
\end{equation}
denote the cross-sectional mean at elapsed day $t_{j}$. Averaging Eq.~\eqref{eq:SM_individual_slope_linear} over trajectory episodes then gives
\begin{equation}
\frac{1}{N}\sum_{k=1}^{N}\hat{s}_{k}=\frac{\sum_{j=1}^{J}(x_{j}-\overline{x})\overline{Y}_{t_{j}}}{\sum_{j=1}^{J}(x_{j}-\overline{x})^{2}}\eqcolon\hat{s}_{\mathrm{mean}}.
\end{equation}
Thus, when the time grid and regression weights are identical, the mean of the trajectory-level OLS slopes equals the OLS slope fitted to the cross-sectional mean curve.
Under the positive-decay convention,
\begin{equation}
\overline{a}_{\mathrm{emp}}\coloneq\frac{1}{N}\sum_{k=1}^{N}\hat{a}_{k}=-\hat{s}_{\mathrm{mean}}.
\end{equation}
The mixture-weighted mean $\overline{a}_{\mathrm{mix}}$ in Eq.~\eqref{eq:SM_gmm_mean} approximates the empirical mean $\overline{a}_{\mathrm{emp}}$ of the fitted slope sample.
For the 2020 calculation,
\begin{equation}
\overline{a}_{\mathrm{emp}}=0.2491189165,\quad\overline{a}_{\mathrm{mix}}=0.2491189165.
\end{equation}
Their absolute difference is below $10^{-12}$ at the numerical precision of the calculation.
The corresponding standard deviations are
\begin{equation}
\sigma_{\mathrm{emp}}=0.3687132593,\quad\sigma_{\mathrm{mix}}=0.3687146154,
\end{equation}
which differ by approximately $1.36\times10^{-6}$.
These comparisons confirm that the fitted Gaussian mixture accurately represents the first two moments of the empirical slope sample.

The coefficient obtained from the ensemble mean in Table~\ref{tab:summary} is different:
\begin{equation}
\hat{a}_{\mathrm{cohort}}=0.277\pm0.002
\end{equation}
for the 2020 cohort.
This coefficient is fitted over the data-adaptive scaling interval $1\leq t\leq\tc$, with $\tc=86$ for 2020.
By contrast, the trajectory-level GMM uses elapsed days $2$--$365$ and therefore summarises relaxation over nearly the full post-selection horizon.
The difference between $0.249119$ and $0.277$ reflects these different regression windows and the finite-time variation of the effective relaxation rate.
The use of $Y_{k,1}$ as the trajectory-level regression reference and $Y_{k,0}$ as the cohort-level mean reference does not itself generate the difference in fitted slopes, because both regressions include an intercept.
The regression window, rather than the additive reference level, is the relevant distinction.
The agreement between $\overline{a}_{\mathrm{emp}}$ and $\overline{a}_{\mathrm{mix}}$ is therefore an internal implementation and density-approximation check.
It is not independent evidence that every trajectory follows a fixed article-specific drift or that the mixture components represent distinct dynamical populations.

\subsection{Scope of the mixture interpretation}\label{SI:gmm_scope}
The fitted coefficients $\hat{a}_{k}$ are effective finite-trajectory summaries.
They can reflect persistent differences in relaxation, stochastic path variation, burst events, temporal modulation, and regression error.
Consequently, the empirical density of $\{\hat{a}_{k}\}$ cannot be identified directly with a distribution of latent article-specific drift parameters.

To see this distinction at the regression level, suppose that a trajectory admits the representation
\begin{equation}
Y_{k,t_{j}}=C_{k}-A_{k}x_{j}+I_{k,t_{j}},
\end{equation}
where $x_{j}=\ln(t_{j}/t_{0})$, $A_{k}$ is a latent decay coefficient, and $I_{k,t_{j}}$ contains stochastic and unmodelled trajectory variation.
Using the common OLS weights from Sec.~\ref{SI:reproduce_a}, the fitted positive-decay coefficient becomes
\begin{equation}\label{eq:SM_effective_decay_decomposition}
\hat{a}_{k}=A_{k}-\frac{\sum_{j=1}^{J}(x_{j}-\overline{x})I_{k,t_{j}}}{\sum_{j=1}^{J}(x_{j}-\overline{x})^{2}}.
\end{equation}
Thus $\hat{a}_{k}$ contains both the latent coefficient $A_{k}$ and the projection of the realised stochastic path onto the log-time regressor.
Even if every trajectory shared the same value of $A_{k}$, finite-path variation in the second term could produce a broad distribution of fitted coefficients.

Under the homogeneous Gaussian reference, $A_{k}=a$ is common to all trajectories and the sampled stochastic vector $\bm{I}_{k}\coloneq\bigl(I_{k,t_{1}},\ldots,I_{k,t_{J}}\bigr)^{\mathsf T}$ is centred multivariate Gaussian with covariance matrix $\bm{\Sigma}_{I}$.
Defining the regression weights
\begin{equation}
w_{j}\coloneq\frac{x_{j}-\overline{x}}{\sum_{\ell=1}^{J}(x_{\ell}-\overline{x})^{2}},
\end{equation}
Eq.~\eqref{eq:SM_effective_decay_decomposition} becomes $\hat{a}_{k}=a-\bm{w}^{\mathsf T}\bm{I}_{k}$.
It follows immediately that
\begin{equation}
\hat{a}_{k}\sim\mathcal{N}\bigl(a,\bm{w}^{\mathsf T}\bm{\Sigma}_{I}\bm{w}\bigr).
\end{equation}
Temporal correlations therefore affect the width of the fitted-slope distribution through $\bm{\Sigma}_{I}$ but not its Gaussian form.
The marked non-Gaussianity of the empirical distribution indicates trajectory-level structure beyond this homogeneous Gaussian reference.
It does not, however, distinguish latent drift heterogeneity from burst events, non-Gaussian fluctuations, temporal modulation, or other forms of model misspecification.
Consequently, neither the width nor the shape of the empirical distribution can by itself be identified with a distribution of latent drift coefficients.

The same distinction appears in the cross-sectional variance.
Consider the continuous-time representation
\begin{equation}
Y_{k}(t)=Y_{k}(t_{0})-A_{k}L(t)+I_{k}(t),\quad L(t)\coloneq\ln\left(\frac{t}{t_{0}}\right).
\end{equation}
Taking the variance across trajectory episodes gives
\begin{align}
\Var_{k}[Y_{k}(t)]={}&\Var_{k}[Y_{k}(t_{0})]+L(t)^{2}\Var_{k}[A_{k}]+\Var_{k}[I_{k}(t)]\notag\\
&{\hspace{-1.75em}}-2L(t)\Cov_{k}[Y_{k}(t_{0}),A_{k}]+2\Cov_{k}[Y_{k}(t_{0}),I_{k}(t)]\notag\\
&{\hspace{-1.75em}}-2L(t)\Cov_{k}[A_{k},I_{k}(t)].
\label{eq:variance_decomposition}
\end{align}
A genuine distribution of latent drifts therefore generates an explicit $L(t)^{2}\Var_{k}[A_{k}]$ contribution, together with covariance terms involving the initial condition and the stochastic component.

If the empirical distribution of $\hat{a}_{k}$ were equated directly with the distribution of $A_{k}$ while all covariance terms were neglected, the resulting $[\ln(t/t_{0})]^{2}$ contribution would not reproduce the observed variance law, which is approximately linear in $\ln t$ over the selected range.
The ensemble-moment analysis does not separately identify $\Var_{k}[A_{k}]$ and the covariance terms in Eq.~\eqref{eq:variance_decomposition}.
A hierarchical model accounting explicitly for finite-trajectory estimation error and stochastic path variation would be required to infer a latent drift distribution.

The Gaussian mixture in Sec.~\ref{SI:gmm} is therefore used only as a flexible approximation to the empirical density of effective decay coefficients.
Its component number, weights, means, and variances are not interpreted as identified microscopic article classes.
The mixture is not inserted into the reduced ensemble model, and no claim is made that drift heterogeneity is the exclusive source of trajectory-level variation.

The substantive empirical conclusion is correspondingly limited but informative.
Broad, non-Gaussian variation in effective trajectory-level relaxation coexists with reproducible logarithmic scaling of the ensemble moments.
This coexistence shows that stable collective scaling is compatible with heterogeneous observed trajectories, without assigning the observed heterogeneity to a unique microscopic mechanism.

\end{document}